# MiBoard: Multiplayer Interactive Board Game

Kyle B. DEMPSEY[a,1], Justin F. BRUNELLE[b], G. Tanner JACKSON[a],
Chutima BOONTHUM[c], Irwin B. LEVINSTEIN[b], and Danielle S. McNAMARA[a]
[a]*University of Memphis, Psychology/Institute for Intelligent Systems*
[b]*Old Dominion University, Computer Science*
[c]*Hampton University, Computer Science*

**Abstract.** Serious games have recently emerged as an avenue for curriculum delivery. Serious games incorporate motivation and entertainment while providing pointed curriculum for the user. This paper presents a serious game, called MiBoard, currently being developed from the iSTART Intelligent Tutoring System. MiBoard incorporates a multiplayer interaction that iSTART was previously unable to provide. This multiplayer interaction produces a wide variation across game trials, while also increasing the repeat playability for users. This paper presents a demonstration of the MiBoard system and the expectations for its application.

**Keywords.** Intelligent Tutoring Systems, Games, Serious Games

## Introduction

Serious games have developed into a serious force in the educational realm. Delivering content to students via entertaining and challenging games has become a legitimate avenue for curriculum developers [1]. Students stand to benefit when developers use games to deliver curriculum because students likely become more engaged, are likely to spend more time on task, and are likely to return for subsequent learning sessions [2]. Serious games must balance entertainment, education, motivation, deliberation, adaptability, and affordability, by identifying which features of the game effectively promote learning while providing interactive entertainment for the student.

Here, we discuss a development project for a serious game called MiBoard. MiBoard is being developed as a serious games extension of the Intelligent Tutoring System, iSTART [3,4]. iSTART (Interactive Strategy Trainer for Active Reading and Thinking) is an automated tutor that teaches users to effectively self-explain texts using reading strategies. iSTART provides curriculum delivery at a one-on-one level without prohibitive cost. However, the current version of iSTART does not adequately address motivational factors for students who are required to use the system over long periods of time (e.g. months). By developing MiBoard, we believe that we are both providing variety in the classroom and improving motivation for long-term use that will result in the users increasing time-on-task (increasing the overall effectiveness of the system), as well as experiencing improved affect towards the system as a whole [5].

---

[1] Corresponding Author

## 1. The Game

MiBoard is an online multiplayer board game that requires players to successfully produce self-explanations as well as identify the strategies used in other players' self-explanations (Comprehension Monitoring, Paraphrasing, Prediction, Elaboration, and Bridging). Within MiBoard, players earn points when a majority of players identify that the same strategy is used within another player's (the *reader's*) explanation. Players can spend these points during the game to change task parameters or activate special "in game" features (e.g., take an extra turn, freeze another player, draw an extra card, etc.). MiBoard does not provide feedback for the players' self-explanations. Instead, players receive feedback from the other players in the game through both modeling of self-explanations as well as through a chat room discussion.

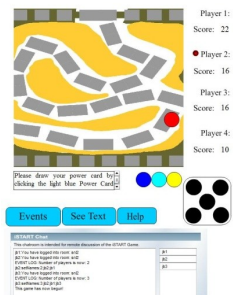
a. Game Screen

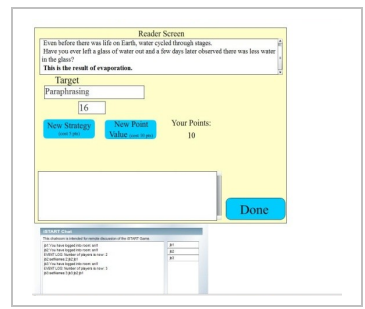
b. Reader Screen

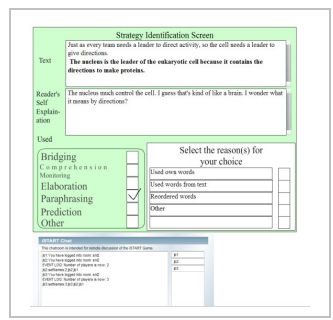
c. Strategy Identification Screen

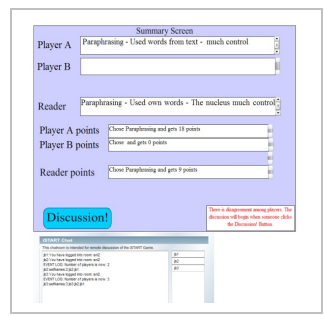
d. Summary Screen

**Figure 1.** MiBoard: screen shots

## 2. Gameplay

MiBoard consists of a computerized representation of a game board that allows players to track their game progress in comparison to the other players. Players are represented on the board as tokens and take rotational turns as either the reader (responsible for producing the self-explanation) or the guesser (responsible for identifying which strategy is used in the self-explanation). Throughout the game, players encounter four different screens: the game screen, the reader screen, the strategy identification screen, and the summary screen (see Figure 1).

*2.1. The Game Screen*

The game screen encompasses the main game board and the choices that can be made during the game portion of the program, incorporating both token movement and game strategy. After completing their turn as the reader, the player first has an opportunity to use a Power Card (if applicable). Power Cards are one of the major motivational tools of MiBoard that allow the player to take a modicum of control over the movement of player tokens around the board. Power Card actions (e.g., freezing an opponent for their next turn) are activated by spending points. Next, the player rolls a die, moves, and draws an Event Card, cueing them to move forward, backward, or draw a Power Card. The *reader* tag passes to the next player at the completion of a turn. This sequence continues until one player reaches the finish, a congratulatory screen is displayed, and players are offered the opportunity to start a new game.

*2.2. The Reader Screen*

Each player takes turn as a *reader* by using an assigned strategy to self-explain a target sentence. The target text appears in context (with the previously presented text) where the reader may review the previous sentences at any time while self-explaining the target sentence. Players may at this point use their accumulated points to alter the potential point value for their self-explanation or alter the given reading strategy. Once a player has read the target text and submitted their self-explanation, the players all enter the strategy identification screen.

*2.3. The Strategy Identification Screen*

After the reader has submitted a self-explanation, the players are shown the strategy identification screen. The players are shown the target text (with context) and the self-explanation and must decide which dominant strategy the reader used in the self-explanation. After indicating which reading strategy they thought was used in the reader's self-explanation, the guessers and reader are moved to the summary screen.

*2.4. The Summary Screen*

At the summary screen, the system displays how everyone voted and awards points if at least half of the players agree on the reading strategy that is used. If the reader is part of the majority, those in the majority are awarded points based upon the points associated with the strategy for that turn, with the guessers receiving half of the points assigned to the category and the reader getting all of the points assigned. If there is a majority that does not include the reader, the guessers are awarded a smaller point total. By reducing the overall points available for an incorrect response by the reader, the players are encouraged to help the reader understand the strategies over the course of the game. If all of the players agree on the reading strategy used, an agreement bonus is awarded, and players are given the option of going directly back to the game screen.
If there is disagreement, the players enter the discussion stage where they are required to resolve their disagreements.
After the discussion phase, players are sent back to the strategy identification screen and are again given the chance to vote on which strategies were included in the self-explanation. After re-voting, players are awarded points for convincing other players to side with their original vote, and the self-explanation is complete. If a player is able to convince another player to choose their strategy, the player is awarded points. This feature allows for players to recoup some points.

## 3. Discussion

A major development of MiBoard is the use of the players as the comprehension check. Currently, there is no automated check against the target reading strategy. Requiring the players to police themselves introduces an interesting aspect that not many games have explored. By having the players discuss their strategy use, but not giving them as many points as they would get for all initially agreeing, the players should be more motivated to understand and apply the reading strategies because doing so will enhance game performance.

When completed, MiBoard will be the end result of the educational and motivational developments in iSTART. MiBoard is a dynamic, competitive environment available to a wide market that requires readers to understand and apply knowledge of reading strategies in order to succeed against other players. By increasing motivation, we expect users to display higher levels of engagement in the system as well as display a stronger desire to initiate a session with MiBoard. Therefore, as players compete in the game, they are likely to engage in the same amount and level of practice as in iSTART, but at the same time will be required to apply their knowledge of strategies by judging others' self-explanations. We expect these aspects of the games to have substantial and meaningful benefits for students' ability to understand challenging texts.

## Acknowledgements

This research was supported in part by the Institute of Educational Sciences (IES R305G020018-02; R305G040046, R305A080589) and National Science Foundation (NSF REC0241144; IIS-0735682). Any opinions, findings, and conclusions or recommendations expressed in this material are those of the authors and do not necessarily reflect the views of the IES or NSF.